# Effect of wall thermal inertia upon transient thermoacoustic dynamics of a swirl-stabilized flame

Giacomo Bonciolini[1], Dominik Ebi[2], Ulrich Doll[2], Markus Weilenmann[1], Nicolas Noiray[1]

[1] *CAPS Laboratory, MAVT department ETH Zürich, Sonneggstrasse 3, 8092, Zurich, Switzerland*
[2] *Laboratory for Thermal Processes and Combustion, Paul Scherrer Institute, 5232, Villigen, Switzerland*

**Abstract**

This paper shows the importance of considering the thermal state of a combustor to investigate or predict its thermoacoustic stability. This aspect is often neglected or regarded as less important than the effect of the operating parameters, such as thermal power or equivalence ratio, but under certain circumstances it can have a dramatic influence on the development of the instabilities. The paper presents experimental results collected from a combustor featuring a lean swirl-stabilized flame exhibiting thermoacoustic instability at some operating conditions. It is shown that this instability is caused by a change of the flame topology that is induced by the progressive increase of the wall temperature with the thermal power. This dependence of the instability on wall temperature leads to inertial effects and hysteresis when the operating condition is changed dynamically. A low-order model of the system reproducing this remarkable dynamics is proposed and validated against the experimental data.

## 1. Introduction

Thermoacoustic instabilities are a recurrent issue in the development phase of combustion systems such as heavy-duty gas turbines, aeronautical or rockets engines [1, 2]. This complex phenomenon arises when the coupling between heat release rate fluctuations and acoustic pressure oscillations becomes constructive in the combustor. Under such conditions, the combustor suffers high-amplitude thermoacoustic limit cycles leading to high-cycle fatigue of the components and potentially to sudden mechanical failure, which must be avoided at all cost. Nowadays, it is still very difficult to predict the occurrence of thermoacoustic instabilities in real engines where the combustion system complexity is markedly increased compared to laboratory layouts [3]. The latter academic configurations nevertheless provide extremely valuable information on the fundamental features of thermoacoustic problems in practical combustors. In these generic configurations, it is indeed easier to systematically investigate the effect of important operating or geometrical parameters on the thermoacoustic stability, e.g. the fuel mass flow distribution between the different burners [4], the inlet temperature and equivalence ratio [5] or the swirler position, which influences the phase

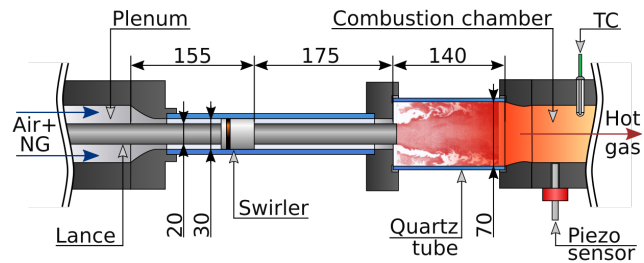

Figure 1: Side view of the laboratory combustor used in this study with key dimensions in mm. The main components are indicated with arrows. All the acoustic measurements presented in this work are recorded by means of a piezoelectric sensor located in the square section of the combustion chamber. The temperature of the ceramic wall is measured with a type-K thermocouple (TC).

of convective disturbances [6].

In this study we present experimental results collected from the modular combustor depicted in fig. 1, in which a perfectly premixed turbulent flame is anchored downstream of an axial swirler. For a given air mass flow, one of the thermoacoustic modes evolves from stable to unstable, and back to stable again as the fuel mass flow is increased. In combustion chambers, it is very usual to encounter non-monotonic thermoacoustic behavior when a control parameter is increased. In many cases, it can be attributed to a monotonic change of the convective time lag separating acoustically-triggered hydrodynamic or compositional perturbations and their interaction with the flame. With respect to the acoustic cycle and for a gradual change of the control parameter, the perturbations may first reach the flame too early, then right on time, and finally too late to lead to a positive Rayleigh criterion. In other cases, this behavior can be due to a non-monotonic change of the amplitude of these acoustically-induced perturbations as function of the control parameter, even though the phasing between acoustic pressure and fluctuating heat release rate remains favorable for a constructive thermoacoustic coupling. When the reacting flow topology does not significantly change, the information associated to these mechanisms is contained in the burner and flame transfer functions, which relate acoustic perturbations and heat release response and which can be used to predict the instabilities using low-order thermoacoustic network [7]. In what follows, it will be shown that in the present configuration the root cause of the non-monotonic thermoacoustic behavior for fuel mass flow variation is linked to a thermally-induced flame topology change, which is very challenging to account for in a thermoacoustic network model. In addition, the present investigation deals with the transient thermoacoustic behavior that is observed when the thermal power is varied at a faster rate than the characteristic time scale that is needed to reach thermal equilibrium between the reactive flow and the combustion chamber wall. This aspect might be relevant to explain peculiar thermoacoustic dynamics that are sometimes observed during cold starts and fast loading of gas turbines. In particular, the effect of thermal inertia and the resulting hysteretic thermoacoustic behavior are analyzed and a simplified model is established using stochastic differential equations.

## 2. Thermoacoustics under stationary conditions

Our generic combustion test rig, whose dimensions and components are shown in fig. 1, was operated at atmospheric pressure with natural gas (NG) and a fixed air mass flow of 18g/s injected at ambient temperature. The thermoacoustic stability is first assessed under stationary conditions, as function of the thermal power $Q$. The NG mass flow was changed from 0.74 to 0.94 g/s. Assuming a lower heating value of 38 MJ/kg, the combustor thermal power $Q$ varied from 28.1 to 35.7 kW, with equivalence ratio being accordingly comprised between 0.71 and 0.90. The NG and the air are injected under choked condition into the plenum where they mix. The mixture then flows through the blades of an axial swirler mounted on a lance and enters the combustion chamber, whose first section is made of a quartz tube. The flame is anchored at the tip of the lance. This configuration ensures the homogeneity of the combusted mixture and a robust anchoring of the flame. At each stationary operating condition, 2 minutes of acoustic pressure in the chamber and integrated flame chemiluminescence were recorded. These data were acquired by means of a water-cooled Kistler piezoelectric sensor type 211B2, and a photomultiplier Hamamatsu H10721-110 equipped with an OH filter. Several thermoacoustic modes can be identified from the spectra displayed in fig. 2a. In the present work, the focus will be on the mode having the eigenfrequency drifting from circa 100 to 120 Hz. Its stability changes with the thermal power. In the upper rows, time traces of the acoustic pressure $p(t)$ at three typical fuel mass flows are presented. In the top panels, the raw data are plotted in gray. The band-pass filtered signal is overlaid in order to highlight the dynamics of the thermoacoustic mode of interest. This filtering bandwidth spans over the shaded region in the panels where the corresponding power spectral densities $S_{pp}$ are displayed. The envelope of the filtered signal, which can be considered as an approximation of the mode amplitude $A(t)$, is also plotted as a darker line. Following [8], it is here assumed that one can reliably write $p(t) \simeq A(t)\cos\left(\omega_0 t + \varphi(t)\right)$ when the mode governs the thermoacoustics dynamics. It is important to note that the amplitude $A$ and the phase $\varphi$ are two quantities that are randomly fluctuating on a time scale sensibly longer than the acoustic period $2\pi/\omega_0$. The driver of these fluctuations is the turbulence, which acts as a non-coherent forcing of the thermoacoustic oscillations [9]. Comparing the three examples, one can see that the modal amplitude is high for the intermediate thermal power $Q$=30.4 kW, while it is low for the other two operating points at 28.1 kW and 34.2 kW. The probability density function (PDF) of the acoustic mode amplitude $P(A)$ is shown in fig. 2b for eleven successive operating conditions from $Q$=28.1 to 35.7 kW. Darker hues of the color scale correspond to states with higher probability density, i.e. amplitudes that are more likely to be visited during stationary operation at that particular power. The evolution of $P(A;Q)$ indicates that the combustion chamber experiences a thermoacoustic instability for the central part of the examined thermal power range. For these operating conditions the amplitude PDF is centered around a large amplitude, suggesting that a stochastically perturbed limit cycle is established. These measurements indicate that the thermoacoustic system experiences two consecutive supercritical Hopf bifurcations when the thermal power is varied. A similar bifurcation diagram is also observed when the thermal power is varied at a constant equivalence ratio. This additional mapping (not presented for the sake of brevity) suggests that the system stability is not significantly influenced by the variation of turbulent flame speed when the equivalence ratio changes.

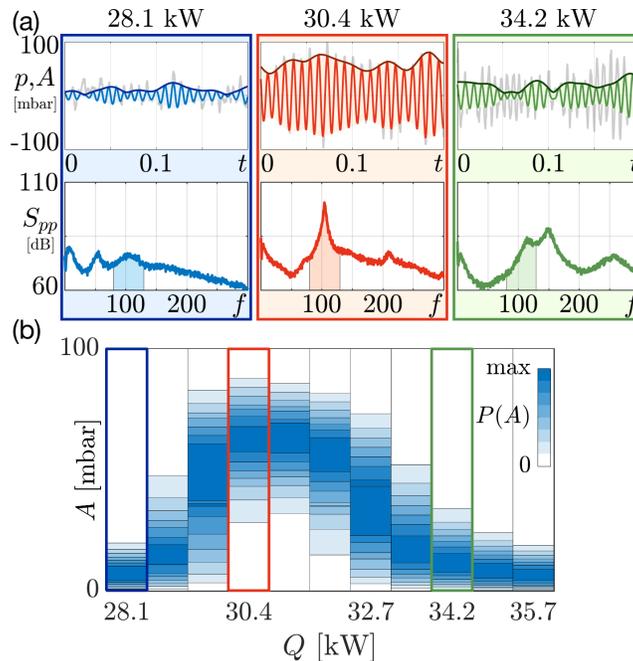

Figure 2: a) Time traces of the acoustic pressure $p$ and its envelope $A$ and corresponding spectra at three example conditions ($Q=$ 28.1; 30.4; 34.2kW). b) $P(A;Q)$, probability density function of $A$ as a function of the thermal power $Q$.

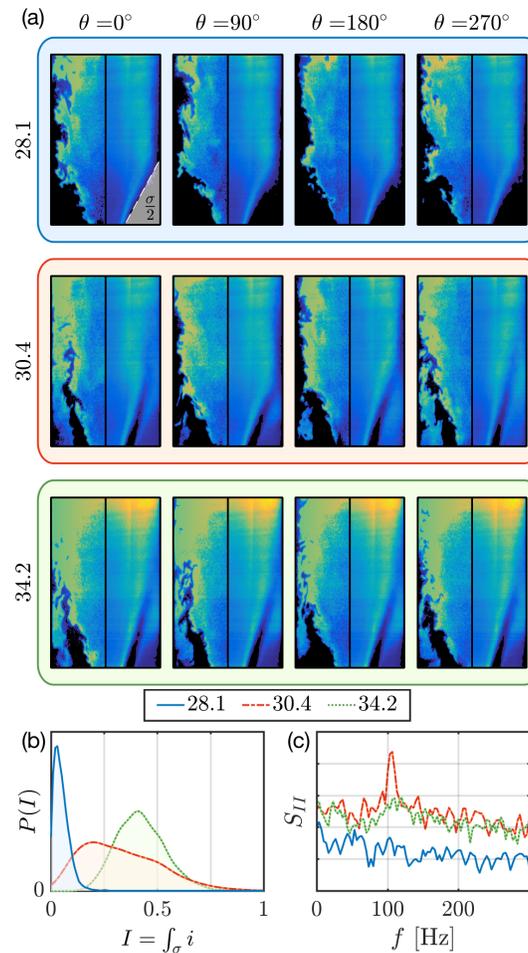

Figure 3: a) Laser Induced Fluorescence images of the flame at the three operating conditions of fig. 2a. From top to bottom: $Q=$ 28.1; 30.4; 34.2kW. In each row, an instantaneous snapshot and the corresponding phase-averaged OH fluorescence at 4 different phases $\theta$ of the acoustic cycle $p(t) = A(t)\cos(\theta)$. b) The PDF $P(I)$ of the OH fluorescence intensity integrated for each of the 3000 snapshots over the two corners of area $\sigma/2$ (the right corner is shown in the top-left image). c) Power spectral density of the integrated OH fluorescence at the three operating conditions.

It is shown in what follows that the thermoacoustically unstable conditions span over the range of fuel mass flow where the mean flame topology is changing. The visualization of the flame front in the central plane of the combustion chamber was obtained with high-speed Planar Laser Induced Fluorescence (PLIF) of the OH radicals at the three operating conditions presented in fig. 2a. The optical measurement system was composed of a high-speed CMOS camera (Photron FASTCAM SA-X2) coupled to a high-speed intensifier (LaVision HS-IRO) equipped with a 100 mm f/2.8 UV lens (Cerco) and a bandpass interference filter (Chroma, transmission >70% at 310 nm, FWHM 10 nm). The laser system was composed of a DPSS pump source (Edgewave IS400-2-L) followed by a dye laser (Radiant Dyes NarrowScan HighRep) and an external frequency-doubling unit. The achieved UV energy per pulse was of 0.5±0.05 mJ and the obtained laser sheet was circa 0.5 mm thick. The images were captured with a frame rate of 2000 fps for 1.5s, and the pixel intensity was corrected with an exponential function that accounts for the progressive light absorption by the OH radicals through the laser sheet path [10]. The resulting images are presented in fig. 3a. Each row corresponds to one of the operating points (same color code as in fig. 2a), and it features four images at four phases $\theta$ of the acoustic cycles $p(t) = A(t)\cos(\theta(t))$. Each image is split in two halves: the left part is an instantaneous PLIF snapshot, and the right part is the phase-averaged flame shape. The flame shape has a key role on the thermoacoustic stability, e.g. [11, 12]. The $Q$=28.1 kW condition is characterized by a thermoacoustically stable V-shaped flame with an outer recirculation zone (ORZ) filled with fresh reactants. The distribution of the ORZ-integrated fluorescence signal $P(I)$, computed from 3000 snapshots with $I_k = \int_\sigma i_k d\sigma$, is plotted in panel b. This distribution has a peak at very low intensity, indicating that the flame does not penetrate the ORZ during the 1.5 s record, which corresponds to about 150 acoustic cycles (one of the 2 ORZ-integration region is shown in the $\theta = 0°$ phase-averaged image). At the other extreme, when $Q$=34.2 kW the flame has an M-shape and hot products are present in the ORZ. The statistic of the ORZ-integrated fluorescence intensity consists of a Gaussian-like distribution with a mean around 40% of the maximum normalized intensity. While these operating regimes are thermoacoustically stable, the intermediate conditions are characterized by intense thermoacoustic limit cycles reaching a RMS acoustic amplitude of about 70 mbar at $Q$=30.4 kW. The flame motion during an acoustic cycle at the condition $Q$=30.4 kW is shown in the second row of fig. 3a. At this condition, the PDF of the ORZ-integrated fluorescence intensity is broad and the power spectrum of $I(t)$, displayed in fig. 3c, features a sharp peak at the acoustic frequency. These two facts indicate that the ORZ is periodically filled with hot products. These cyclic transitions from V to M shape are accompanied with intense coherent heat release rate fluctuations, as already reported, for instance, in [13, 14]. These heat release rate oscillations constitute a significant acoustic source, leading to the thermoacoustic instability.

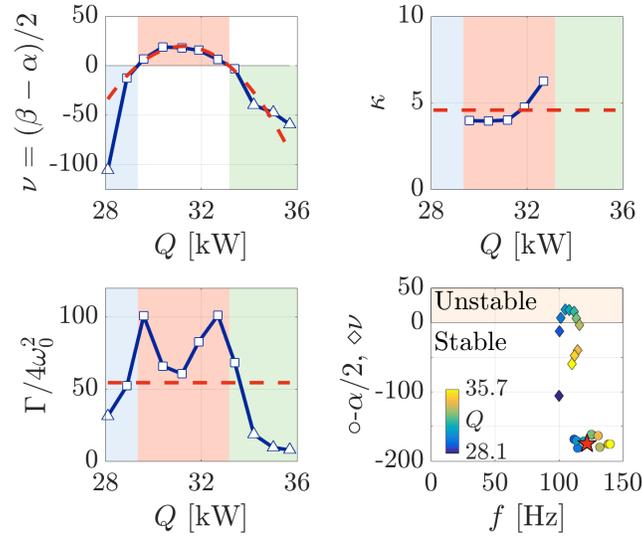

Figure 4: Identification of the low-order model parameters. The linear growth rate $\nu$, the cubic saturation constant $\kappa$ and the noise intensity $\Gamma$ are identified with the output-only methods proposed in [8]. The background colors of these three panels indicate the linear stability of the system, as deduced from the sign of $\nu$. In the panel on the bottom right, the linear damping $\alpha$ identified with the input-output method based on acoustic and photomultiplier signals proposed in [15]. The symbols $\circ$ represent the poles of the acoustic system with a "thermoacoustically inactive" flame, having real part $-\alpha/2$, at the eleven analyzed operating conditions. The thermoacoustic feedback induces a significant shift of these poles, which are moved to $\diamond$, located at $(f_{\text{peak}}(Q), \nu(Q))$. The dashed lines and the star indicate the values adopted for $\nu(Q)$, $\kappa$, $\Gamma$, $\alpha$ and $\omega_0 = 2\pi f_0$ in the simulations performed in section 5.

## 3. Identification of governing parameters

It is interesting to introduce a low-order phenomenological model of the thermoacoustic dynamics with a reduced number of parameters that can be identified from the experimental records: the Van der Pol (VDP) oscillator. This model has been already used in a thermoacoustic context, for instance in [8, 16, 17, 18]. In the present case, the turbulence in the combustion chamber acts as a stochastic forcing on the VDP oscillator, whose state $p$ represents the local acoustic pressure resulting from the self-sustained oscillation of the dominant thermoacoustic eigenmode:

$$\ddot{p} + \omega_0^2 p = [2\nu - \kappa p^2]\dot{p} + \xi. \tag{1}$$

In this stochastic differential equation, $\nu = (\beta - \alpha)/2$ is the linear growth rate, whose sign determines the linear stability of the oscillator. This coefficient is determined by the linear gain $\beta$ and the linear damping $\alpha$. The other model parameters are $\kappa$, the non-linear saturation coefficient, and $\xi$, a white noise source of intensity $\Gamma$ [9]. To identify the parameters $\nu$, $\kappa$, $\Gamma$ it is possible to use the output-only identification strategies given in [8]. In this case, the input of the system is the stochastic forcing exerted by turbulence, which cannot be directly measured. These identification methods require only the output of the system, i.e. the pressure signal $p$ (or its envelope $A$). The identified parameters are presented in the panels of fig. 4. More specifically, method 4 from [8] was applied for the operating conditions where the thermoacoustic mode around 100 Hz dominates the thermoacoustic dynamics ($\square$). This method relies on the extraction of the drift and diffusion coefficients of the Fokker-Planck equation describing the evolution in time of the amplitude PDF $P(A;t)$. At low and high power, the parameters are not identified in a robust way using this method because neighboring eigenmodes significantly contribute to the thermoacoustic dynamics and cannot be filtered out. For these points, method 1 from [8] was used ($\triangle$). This method fits a Lorentzian function on the eigenmode peak in $S_{pp}$ and the identification of the parameters is less sensitive to the presence of neighboring eigenmodes. However, this method can only be applied to linearly stable conditions. At low and high power $\nu$ is negative and therefore the system is linearly stable, while for intermediate powers $\nu > 0$ and the system exhibits a limit cycle. The evolution of the identified $\nu$ is approximated with a parabola (dashed line on the plot). Making use of this fit, it is possible to infer the stability borders, which are marked with a change in the background color in the diagrams. The identified noise intensity exhibits a peculiar dependence on the thermal power of the combustor $Q$ and significantly rises at intermediate power, which may be related to the hydrodynamically unstable nature of the flame at these conditions. Indeed, the transition from V- to M- flame shape may be accompanied with a more intense non-coherent component of the heat release rate fluctuations. On the contrary, the non-linear saturation coefficient $\kappa$ stays almost constant. Finally the input-output identification method described in [15] was applied to identify the linear damping $\alpha$. These results are summarized in the bottom-right panel of fig. 4, where the identified linear damping and growth rates are shown ($\circ$ and $\diamond$ respectively).

## 4. Thermoacoustics during transient conditions

The transitory dynamics is now investigated. Instead of operating the system at a stationary condition, the fuel mass flow was increased at a fast rate. The results are presented in fig. 5, where a sequence of PLIF snapshots recorded during the ramp is shown together with the acquired signals. On each panels of fig. 5 there are 8 dots indicating the instants at which the 8 PLIF snapshots were taken. The signal recorded from the mass flow controller is plotted in the first panel. The convective time from the fuel injector to the combustion chamber is less than 0.1s. The ramp lasts 1 s, and the background colors represent the stability regions obtained from the analysis at stationary conditions. The two vertical black dashed lines indicate the instants at which $Q$ crosses the borders of those stability regions. The second panel shows the amplitude evolution $A(t)$ during the ramp and, on the background, the PDF $P(A;t)$ obtained repeating this experiment 50 times. The transition to the limit cycle is delayed; the acoustic oscillations have a small amplitude even when the thermal power $Q$ corresponds to an unstable point in the stationary case (snapshots 2 and 3). In addition, the limit cycle persists even at high power, where in the stationary case the system should be linearly stable (snapshots 5 to 8). Therefore it can be concluded that the thermal power is not the significant bifurcation parameter in this system, as it is not possible to establish a bijection between its value and the system dynamics. In the stationary case, any thermal power corresponds to a steady wall temperature distribution. In the ramped case, the temperature rises up progressively, with a relaxation time that depends on the components material and on the convective heat transfer. The third panel shows the surface temperature of the ceramic wall downstream of the flame measured with a type-K thermocouple (its position is indicated in fig. 1). One can observe how the temperature is still rising at the end of the ramp; the thermal equilibrium is not yet reached. Since the flame is highly confined in this combustion chamber, having an area ratio with the burner exit of 5.4, one can expect a strong flame-wall interaction [19]. As demonstrated in previous studies [20, 21] the wall temperature can induce a flame topology transition, which can change the thermoacoustic stability of the system [22, 23, 24]. In [25], the thermal transient of the annular chamber wall for fixed air and fuel mass flows is accompanied with a drift of the thermoacoustic dynamics. In the present case, a similar phenomenon is observed when the fuel mass flow is increased. Figure 5, which displays the fluorescence signal integrated over the ORZ, indicates that the flame does not penetrate the ORZ until the temperature has significantly risen compared to the thermal equilibrium temperature of the initial condition. This can also be observed in the first 3 PLIF snapshots of fig. 5, which all show a V-shaped flame. Afterwards, the mean intensity of the ORZ-integrated fluorescence suddenly increases, which co-occurs with large amplitude oscillations at the acoustic frequency; the walls are hotter and favor periodic switches from V to M (see the rest of the LIF sequence). The transition to a stable M flame, and its corresponding thermoacoustic stabilization, happen after a longer time (not within the acquisition time limit of the high speed camera) when the equilibrium temperature at the high power condition is reached. At the specific location where the thermocouple is placed, the equilibrium temperature of the wall is ca. 1250 °C, and it is reached in approximately 30 s. One can safely assume that the quartz walls around the ORZ exhibit a similar thermal transient, up to the point when the flame switches to the M shape and sits in this area, increasing the local wall temperature and further promoting the M shape.

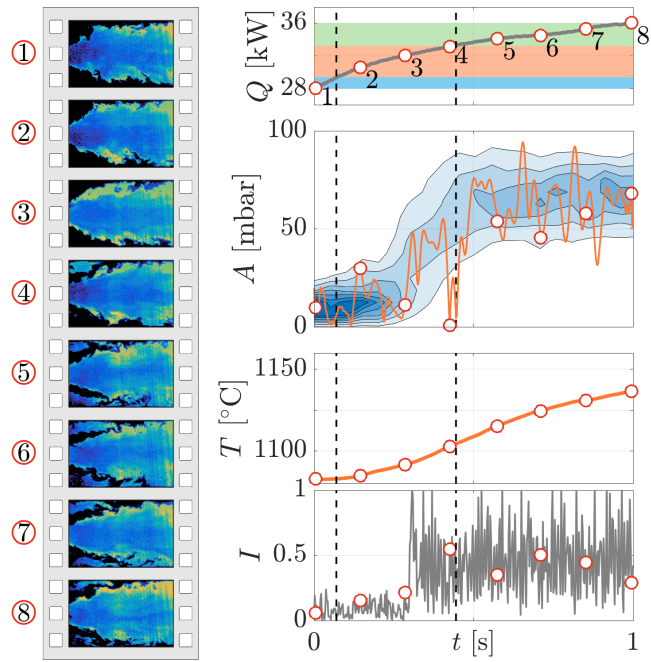

Figure 5: Left: PLIF of the flame during the fuel mass flow ramp showing the delay in the transition from V to flapping V-M. Right: Time traces relative to the ramp experiment. The eight dots correspond to the instants at which the PLIF flame pictures were recorded. From top to bottom: fuel mass flow rate signal converted to the equivalent thermal power; acoustic amplitude $A(t)$ during the ramp, superimposed on the PDF $P(A; t)$ obtained by repeating the ramping 50 times; Surface temperature of the ceramic wall downstream of the flame; Integrated fluorescence intensity $I$ in the combustion chamber corners (compare to fig. 3a-b), showing the transition from V to flapping V-M flame.

Another test was performed in order to capture the complete thermal and thermoacoustic relaxation. Instead of repeated ramps starting from a stationary condition at $Q$=28.1 kW, a 4-step cycle of the thermal power was repeated 50 times. The power is first ramped from the minimum $Q$=28.1 kW to the maximum $Q$=35.7 kW in circa 1.5 s and it is then kept constant at its maximum value for 13.5s. Then $Q$ is ramped down to the minimum (again in circa 1.5 s) and finally kept constant at this value for another 13.5 s. Since the thermal transient is only partially completed at each cycle, the average temperature of the chamber walls is higher than in the previous case. This results in a reduced delay of the flame topology transition. Figure 6 presents the PDF $P(A; t)$ of the 50 ramps up and down extracted from this cyclic measurement. It can be noted that during the ramp up of $Q$, the thermoacoustic instability occurs earlier than in the ramp down case. During the ramping down, the mean of the acoustic envelope statistic rises fast but decreases at a slower rate. This is because when the flame starts as an M-shaped the walls around the ORZ are hot and promote spontaneous flame penetrations in this zone, which sustain the instability and slow down the wall cooling process.

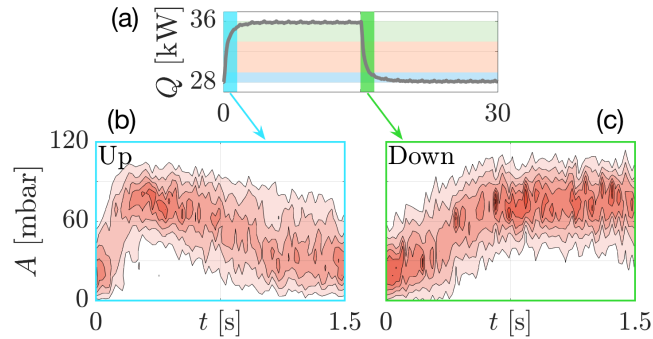

Figure 6: a) One of the fifty power cycles. b-c) PDFs $P(A; t)$ of the ramps up and down extracted from the cyclic experiment, showing the thermal hysteresis.

## 5. Low-order model

A parametric version of the model (1) with time-varying linear growth rate $\nu(t)$ is now considered to reproduce the observations presented in the previous section. The other coefficients of the model ($\alpha$, $\kappa$, $\Gamma$ and $\omega_0$) are set to the mean values identified by processing experimental data recorded under stationary conditions (see the dashed lines and the star in fig. 4). We assume that, i) the equilibrium temperature $T_\infty$ of the wall bounding the ORZ is linearly linked to the thermal power $Q$, ii) for transient conditions, the temperature of the wall $T(t)$ relaxes to $T_\infty(t)$ according to a simple first order differential equation with time constant $\tau_T$ that is determined by the system configuration, the components materials and the operating conditions, and iii) the *instantaneous* linear growth rate $\nu(t)$ varies with the *instantaneous* temperature $T(t)$ of the wall just as in *stationary* conditions.

In the present experiment, $Q$ is progressively ramped up or ramped down as $Q(t) = Q_1 + (Q_2 - Q_1)\tanh(t/\tau_R)$, where $Q_1$ and $Q_2$ are the initial and final values of the thermal power, and $\tau_R$ is the characteristic time of the fuel valve opening. Considering assumption i), one obtains $T_\infty(t) = T_1 + (T_2 - T_1)\tanh(t/\tau_R)$. Together with assumption ii), this time evolution of the equilibrium temperature leads to the ODE governing the instantaneous temperature of the wall

$$\dot{T}(t) = [T_\infty(t) - T(t)]/\tau_T$$

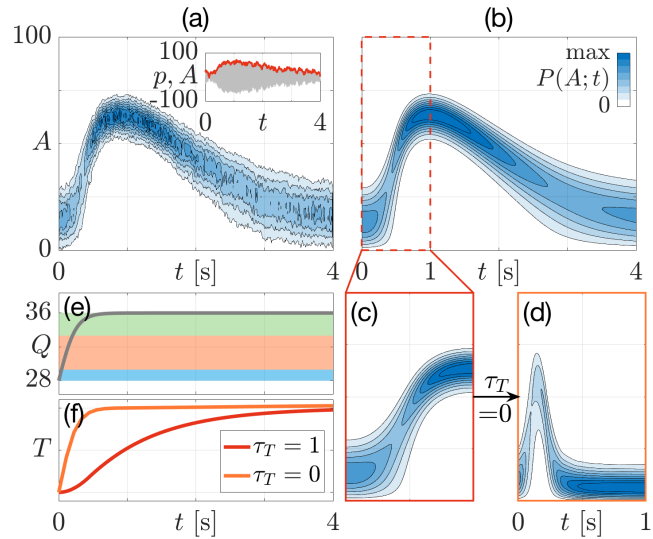

Figure 7: Simulations of the low order model. a) A single realization, with output $p(t) = A(t)\cos(\theta)$, and the ensemble PDF $P(A;t)$ obtained via Simulink® simulations. b) Numerical solution of the Fokker-Planck Equation for $P(A;t)$. c) Detail of the first second of solution compared with the case d) where the thermal relaxation equation is deactivated. e-f) Evolution in time of the control parameter $Q$ and of $T$, for the two simulated cases.

Under stationary conditions, the linear growth rate $\nu$ exhibits a quadratic-like dependence on $Q$ or, following i), on $T_\infty$, which is presented in fig. 4). According to assumption iii), one can therefore write that $\nu(t) = a_\nu T(t)^2 + b_\nu T(t) + c_\nu$. This last assumption is based on the fact that temperature is the parameter actually governing the bifurcation.

Put together, these elements enables one to phenomenologically describe the transient thermoacoustic dynamics, which is done here using Simulink®. The ORZ wall temperature distribution is not known. Therefore, the coefficients for linear relationship between $T_\infty$ and $Q$ are arbitrarily set, and the time constant $\tau_T$ is adjusted in order to match the observed relaxation time. The results are presented in fig. 7a. The PDF $P(A;t)$ obtained from 500 realizations of the process (one example in the inset) is presented in the bottom row. It is also possible to directly obtain $P(A;t)$ by solving the Fokker-Planck equation (FPE) [18]:

$$\partial_t P = -\partial_A [\mathcal{F}(A,t)P] + (\Gamma/4\omega_0^2)\partial_{AA} P, \quad (2)$$

where $\mathcal{F}(A,t) = A[\nu(t) - (\kappa/8)A^2] + \Gamma/(4\omega_0^2 A)$. The initial condition for this equation is the stationary PDF $P_\infty(A)$ at the initial state, which is a function of the system's parameters and of the initial growth rate, and which can be derived analytically[1]. The solution of this FPE is plotted in fig. 7b. The two approaches are in perfect agreement and, looking at the detail shown in panel (c), they reproduce the transient experimental results presented in fig. 5. An additional solution of the FPE was computed changing $\tau_T$ to 0. The result is presented in panel (d). As displayed in panel (f), this case has no thermal-like inertia and therefore the system behaves according to the linear stability of the stationary case.

---

[1] $P_\infty(A, t=0) = \mathcal{N} A \exp\left[4\omega_0^2/\Gamma(\nu_0 A^2/2 - \kappa A^4/32)\right]$, where $\nu_0 = \nu(t=0)$ and $\mathcal{N}$ is a normalization constant. See [9] for more details.

## 6. Conclusions

This paper presents an experimental investigation of the transient thermoacoustic behavior of a lab-scale combustor with swirl-stabilized turbulent flame when the fuel mass flow is ramped up. It is shown that changing the thermal load affects both the flame topology and the thermoacoustic stability. Under stationary conditions at low and at high natural gas mass flow the thermoacoustic state is linearly stable, but the flame exhibits a V shape in the former case and an M shape in the latter. This is because the higher wall temperature, which is associated with the richer condition, allows the outer recirculation zone to be filled with hot products and a flame branch to self-sustain at the outer rim of the burner outlet. However, the system is thermoacoustically unstable in the intermediate range of fuel mass flow and the corresponding large amplitude heat release rate fluctuations result from the periodic switch between M and V shape. It is shown that under transient load, inertial thermal effects govern the thermoacoustic stability. This non-trivial thermoacoustic transient is reproduced with a phenomenological model, featuring a relaxation-type equation for the fluid-structure heat transfer that prescribes the evolution in time of the bifurcation parameter. This work demonstrates that it may be crucial to account for this complex dynamic heat-transfer mechanism in thermoacoustic network models if one wants to reliably predict thermoacoustic transients during change of operating conditions.

## Acknowledgments

This research is supported by the Swiss National Science Foundation under Grant 160579.